\documentclass[10pt,leqno]{article}
\usepackage{indentfirst,csquotes}

\usepackage{amssymb,amsthm,amsmath}
\usepackage{xcolor,paralist,hyperref,titlesec,fancyhdr,etoolbox}

\hypersetup{ colorlinks=true, linkcolor=black, filecolor=black, urlcolor=black }

\usepackage{lipsum}
\usepackage{epsfig}
\usepackage{algorithm}
\usepackage{algpseudocode}
\usepackage{caption}
\usepackage{subcaption}
\usepackage{graphicx}%
\usepackage{multirow}%
\usepackage{amsmath,amssymb,amsfonts}%
\usepackage{amsthm}%
\usepackage{mathrsfs}%
\usepackage[title]{appendix}%
\usepackage{xcolor}%
\usepackage{textcomp}%
\usepackage{manyfoot}%
\usepackage{booktabs}%
\usepackage{algorithm}%
\usepackage{algorithmicx}%
\usepackage{algpseudocode}%
\usepackage{listings}%
\usepackage{caption}%
\usepackage{lineno}%
\usepackage{xurl}%

\begin{document}



\title{Unveiling Molecular Moieties through Hierarchical Grad-CAM Graph Explainability}
\author{Salvatore Contino$^{1, *}$, Paolo Sortino$^{1}$, Maria Rita Gulotta$^{2}$, \\Ugo Perricone$^{2, *}$, Roberto Pirrone$^{1}$}

\maketitle
\let\thefootnote\relax
\footnotetext{$^{*}$corrisponding author salvatore.contino01@unipa.it (Salvatore Contino)} 
\footnotetext{$^{1}$Department of Engineering, University of Palermo, 90128 Palermo, Italy} 
\footnotetext{$^{2}$Molecular Informatics Group,Fondazione Ri.MED, 90128 Palermo, Italy}


\abstract{\textbf{Background:} Virtual Screening (VS) has become an essential tool in drug discovery, enabling the rapid and cost-effective identification of potential bioactive molecules. Among recent advancements, Graph Neural Networks (GNNs) have gained prominence for their ability to model complex molecular structures using graph-based representations. However, the integration of explainable methods to elucidate the specific contributions of molecular substructures to biological activity remains a significant challenge. This limitation hampers both the interpretability of predictive models and the rational design of novel therapeutics.\\
\textbf{Results:} 
We trained 20 GNN models on a dataset of small molecules with the goal of predicting their activity on 20 distinct protein targets from the Kinase family. These classifiers achieved state-of-the-art performance in virtual screening tasks, demonstrating high accuracy and robustness on different targets. Building upon these models, we implemented the Hierarchical Grad-CAM graph Explainer (HGE) framework, enabling an in-depth analysis of the molecular moieties driving protein-ligand binding stabilization. HGE exploits Grad-CAM explanations at the atom, ring, and whole-molecule levels, leveraging the message-passing mechanism to highlight the most relevant chemical moieties. Validation against experimental data from the literature confirmed the ability of the explainer to recognize a molecular pattern of drugs and correctly annotate them to the known target.
\textbf{Conclusion:} Our approach may represent a valid support to shorten both the screening and the hit discovery process. Detailed knowledge of the molecular substructures that play a role in the binding process can help the computational chemist to gain insights into the structure optimization, as well as in drug repurposing tasks.}

\linespread{2.0}

\section{Background}\label{sec1}

The development of a new drug is a long and time-consuming process that, despite advances in technology and computer applications, is fragmented into several steps toward the identification of a lead compound from \textit{in vivo} models. During the last decade, the use of deep learning has provided support to Virtual Screening (VS) campaigns as an effective predictive means to select hit compounds through the use of massive data in the training model phase \cite{Kimber_2021, carpenter2018, Unterthiner2015DeepLA, bahi2018, ZHOU202057}. In fact, being able to correctly prioritize active molecules during the screening steps speeds up the discovery process. The classical approaches used for virtual screening are not always efficient, being affected, for example, by the use of specific force fields \cite{Sun_Huggins_2022, Lin_MacKerell_2019} or the quality of structural data available, thus affecting the whole discovery process from hit identification to hit optimization steps towards the lead compound.
Hit discovery is the very first step of a drug discovery workflow, and chemists use it as the starting point to rationally optimize molecules up to the preclinical candidate. The molecular optimization steps involve different techniques such as hit expansion \cite{Abdolmaleki_2017}, (bio)isosteric replacement \cite{Dick_Cocklin_2020} or a combination of them \cite{Keseru_Makara_2006}. 
Thus, optimization requires an understanding of the relevant chemical features within the small molecule. This step is usually handled by computational chemists who recognize the molecule's activity and consciously act to improve it in accordance with their knowledge and expertise in the field. 

In this context, Graph Neural Networks (GNNs) \cite{scarselligori2009, kipf2016semi} have emerged as a powerful tool in computational drug discovery, offering unique capabilities to model molecular structures \cite{Bongini_Bianchini_Scarselli_2021, Xiong_Xiong_Chen_Jiang_Zheng_2021, Wieder_Kohlbacher_Kuenemann_Garon_Ducrot_Seidel_Langer_2020, Kearnes_McCloskey_2016}. 
Unlike traditional machine learning models that rely on precomputed molecular descriptors, GNNs operate directly on graph representations of molecules, where atoms are nodes and bonds are edges \cite{Jiang_Wu_Hsieh_Chen_Liao_Wang_Shen_Cao_Wu_Hou_2021}. 
This allows GNNs to effectively capture the intrinsic connectivity and topology of molecular structures.

Through iterative message passing mechanisms \cite{gilmer2017neural}, GNNs propagate information across the graph, thus enabling the model to learn complex relationships between the atom neighborhoods and the molecular properties. This data-driven approach reduces the dependence on hand-crafted features, instead leveraging the model's ability to extract relevant patterns directly from the molecular graph. 
In addition, GNN architectures can be tailored to specific tasks, such as predicting molecular properties, bioactivity, or drug-target interactions by integrating domain-specific information through customized layers and loss functions \cite{Zeng_Tu_Liu_2022, yang2019, Chen_Liu_Wu_2019, Wang_Du_Song_2022}.

Even if the graph representation of a molecule is self-explainable, the precise relation between the relevant molecular properties and the network prediction is not, and it suffers from the same issues observed in classical Convolutional Neural Networks (CNN) devoted to images. As a consequence, graph-oriented eXplainable Artificial Intelligence (XAI) techniques have emerged in the Drug Discovery domain. 
A very recent work by Proietti et al. \cite{Proietti_2024}
addresses the \emph{Concept Withening} (CW) technique to investigate the decision process of a Graph Convolutional Neural Network (GCNN) aimed at discovering relevant molecular properties on a plethora of benchmark data sets. CW has been presented in \cite{chen2020concept}: it is a representation learning technique aimed at aligning the axes in the latent space close to some known ``concepts'' (i.e. group of features) in the data space. 
Proietti and his colleagues used GNNExplainer \cite{Ying_Bourgeois_2019} to learn the contributions of concepts in the predictions of a QSAR model. GNNExplainer is an explainer capable of identifying compact subgraphs and small subsets of features through the use of Mutual Information ($MI$) allowing the GNNExplainer to be formulated as: 

\begin{equation}
\max _{G_S} M I\left(Y,\left(G_S, X_S\right)\right)=H(Y)-H\left(Y \mid G=G_S, X=X_S\right) 
\end{equation}

\vspace{2mm}
The $MI$ term measures how much information about the predicted label distribution $Y$ is gained by knowing both a suitable subgraph $G_S$ and a subset of the features $X_S$. By maximizing $MI$, the model seeks the subgraph $G_S$ and the features subset $X_S$ that provide the most information about $Y$. This is achieved by minimizing the conditional entropy $H\left(Y \mid G=G_S, X=X_S\right)$, which indicates that knowing the chosen subgraph and features reduces the uncertainty in predicting $Y$ as much as possible.
In addition to the GNNExplainer, other XAI frameworks have been adapted to explain the information obtained from GNNs. Specifically, the two most widely used methods are SHapley Additive exPlanation (SHAP) and Gradient-weighted Class Activation Mapping (Grad-CAM). SHAP aims to explain the contribution of individual nodes and edges (features) in a GNN to the final prediction for a molecule. To assess the importance of either node or an edge, SHAP creates a ``coalition'' centered around that specific feature. This coalition includes the node/edge itself and all its descendant nodes and edges in the graph. 
SHAP then compares the prediction of the model for the molecule with and without the entire coalition. This difference reflects the impact of the target node/edge and its connected features on the prediction. 
By analyzing SHAP values for different coalitions centered on individual nodes and edges, researchers gain insight into the relative importance of each feature for the GNN's prediction.
In the equation \ref{eq:shap}, $g$ is the explanation model, $z^{\prime} \in{0,1}^M$ is the coalition vector, $M$ is the maximum coalition size and $\phi_j \in \mathbb{R}$ is the $j$-th feature attribute. 

\begin{equation}
    g\left(z^{\prime}\right)=\phi_0+\sum_{j=1}^M \phi_j z_j^{\prime}
\label{eq:shap}
\end{equation}

The equation shows how the SHAP framework breaks down model predictions into contributions from single features or feature combinations, giving each a weight determined by its relative importance. In \cite{Rodríguez_Pérez_Bajorath_2020_1, Rodríguez_Pérez_Bajorath_2020_2}, SHAP was used to determine the significance of individual features; however, it did not provide insight into their pharmacophoric characteristics.

On the other hand, Grad-CAM \cite{Selvaraju_2020} is an explainability technique widely used in image classification through the use of CNNs. It generates a heat map for each of the classes to highlight the parts of an image that contribute to the recognition of that class. At each layer where Grad-CAM is applied, the role of each feature map in explaining the class is weighted using the gradients backpropagated to the layer itself. In~\cite{Pope2019ExplainabilityMF}, a graph-based adaptation of Grad-CAM is introduced, which is called GCNN-Explainability. This method incorporates a Grad-CAM layer in the final stage of the GCNN architecture to identify substructures, that play a significant role in individual classification, thus highlighting features that are representative of a specific class.
The application of the Grad-CAM technique to graph-based models, particularly in the context of molecular graphs, offers significant advantages. By visualizing the most relevant graph components for a classification task,
these methods can shed light on the key pharmacophoric elements within a molecule. This not only aids in understanding the decision-making process of the model, but also provides valuable insights for drug discovery and molecular design, by focusing attention on the structural components that drive biological activity. 

In this paper, we present the Hierarchical Grad-CAM graph Explainer (HGE) framework, which is designed to identify the most relevant molecular substructures, that influence activity
prediction in two data sets of small molecules on twenty different Kinase protein targets. HGE leverages Grad-CAM to obtain local explanations at atom, ring, and the entire molecule level, and then it combines local explainers to provide information on the most relevant molecular moieties.
To prove the effectiveness of HGE we built twenty GCNN classifiers, each dedicated to a specific protein target. These models were trained to optimize classification performance on a single target. In each classification task, the ground-truth activity labels were sourced from ChEMBL.
The proposed HGE framework extends Grad-CAM by providing a deeper analysis of the relevant sub-graphs directly linked to the chemical functionality of the identified moieties, focusing on the
substructures directly involved in predicting an active molecule, rather than providing a global explanation of the classifier behavior.
These substructures or portions of them are expected to play a crucial role in binding to the target. As a consequence, we used a hierarchy of explainers placed at different layers to catch information directly at the atom, ring, and whole molecule level by leveraging the message passing mechanism that is used in GNNs. 

The analysis carried out on a set of already approved drugs demonstrates that trained neural networks can successfully identify inhibitors for each of the 20 targets with a high success rate. 
HGE proved to be very effective in detecting the same common substructures in different molecules that are active on the same target. Conversely, HGE selected the proper diverse substructures for the same molecule when its activity is investigated versus different targets.

Next sections of the paper are arranged as follows: Section \ref{sec2} describes the creation of the two data sets used for training the networks, their structure, and the development of the explainability algorithm. Section \ref{sec3} reports the classification performance analysis of the implemented GNNs with respect to different metrics, and the results obtained from HGE, along with a discussion of the interpretability of these results. Finally, Section \ref{sec5} sets the groundwork for future studies.

\section{Methods}\label{sec2}
The main objective of this work is to develop a method to identify which molecular substructures play a major role in characterizing a molecule as putatively active against a protein target. The analysis of the results provided by HGE may guide medicinal chemists in the design of novel therapeutics for a certain target, focusing the efforts on specific substructures by generating all the relevant decorations to increase the potency of a molecule or build novel putative lead compounds.
For this purpose, we chose one of the major investigated protein families as drug targets \cite{Santos_2017}, i.e. Kinase proteins, to build and test our HGE model. Kinases are essential regulators of numerous cellular processes, including signal transduction, cell cycle progression, and apoptosis, making them critical targets in cancer research and therapeutic development. Moreover, a massive amount of data is available from the literature, making kinases suitable for our work. However, the study described in this work was narrowed to a set of 20 kinases, while the choice of these proteins was described in \cite{mendolia_2022}. These kinases are listed in Table S1 in the supplementary file S1. 

\subsection{Data preparation}\label{data_preparation}
All the experiments were carried out using two data sets: EMBER and ChEMBL\_over. EMBER was curated by some of the authors~\cite{mendolia_2022}, and it was already used in previous research. Data were appropriately encoded in the graph format for the purposes of this work. ChEMBL\_over has been designed purposely for this study with the aim of stressing the classification performance in case of extreme unbalanced data. The data set was derived from the ChEMBL database \cite{10.1093/nar/gkad1004} and enriched by oversampling inactive compounds for each target.
The following sections will provide a detailed explanation of the creation processes for both data sets.

\subsubsection{Data set 1: EMBER}
The first data set used in this paper has been proposed by some of the authors in the work by Mendolia et al. \cite{mendolia_2022}. 
The input molecular graphs used by our GCNNs were generated from the canonical SMILES representation. Unlike the other molecular descriptors used in classical approaches for Virtual Screening, molecular graphs account for chemical information both in each atom and its chemical neighborhood.
Canonical SMILES were recovered from the CHEMBL24.1 database \cite{10.1093/nar/gkad1004} looking for all small molecules whose activity on the target protein is referenced by the values of IC$_{50}$, K$_i$, and K$_d$. We used a fixed threshold activity value on each target in order to generate the molecular graphs through the use of the RDKit library.

Each node represents an atom as a feature vector and nine features were selected in the proposed representation:

\begin{itemize}[nosep]
    \item Atomic number
    \item Degree\footnote{An atom's degree is defined as the number of directly bound neighbors. The degree is independent of bond orders, however it depends \cite{Landrum_2010}}
    \item Formal Charge
    \item Hybridization
    \item Aromaticity
    \item Total number of hydrogen atoms
    \item Number of radical electrons
    \item Information on aromaticity
    \item Chirality
\end{itemize}
\vspace{2mm}

The set of atom features was later augmented with the 3D coordinates of each atom in active configuration, for the sake of training the single-class GCNN architecture used for the hierarchical explainability analysis with respect to each Kinase protein target. 
These coordinates were calculated by minimizing the conformation of the molecule and creating the correct protonation state using Ligprep \cite{schroeding_2023} with $pH=7.4$.
In each molecular graph, the edge information encodes the type of bond (double or triple bond) and whether the bond is within a cycle. The size of each molecular graph had several nodes and arcs which vary according to the size of the molecule.

As reported in the paper by Mendolia et. al, inactive compounds in the EMBER data set were devised using the opposite of the concept of similarity, that is, \textit{dissimilarity}. A compound is labeled as inactive when it is ``dissimilar" to the target: its similarity coefficient to the active compounds is less than $0.1$.
At first, several inactive compounds were discovered with respect to all the targets on ChEMBL. Then each class was further enriched with compounds that are inactive specifically on the target even if they are active on other targets.
The final data set consists of 89373 small molecules and has a 1:100 active/inactive ratio for the smallest class, which refers to the CLK2 target protein. This ratio allowed us to deeply assess the model generalization abilities as it has more or less the same size as a real-world screening task. Training, validation, and test sets were created using a ratio 80\%:10\%:10\% for each target to minimize comparison bias.

\subsubsection{Data set 2: ChEMBL\_over}\label{dataset_2}
A second data set biased towards very unbalanced classes has been developed purposely for this work, that is ChEMBL\_over.
The data set was initially created through direct download using the ChEBML API. Specifically, the organism was set to \texttt{human}, and the relevant \texttt{target\_chembl\_id} and \texttt{target\_type} for the protein were identified. 
Once these initial parameters were defined, the query was constructed by selecting the proper values of $IC_{50}$, $K_d$, $K_i$, $K_{dapp}$, \textit{Inhibition} and \textit{Activity}.

Each small molecule downloaded using the API was labeled based on the threshold values given in the literature. 
The molecules obtained in this way have a very low active/inactive ratio. For this reason, the ChEMBL\_over data set was increased using an oversampling algorithm. This is a very common practice in machine learning to compensate for unbalanced data, and it has been adopted in recent years also in the chemical domain  \cite{Datta_2024, Hareharen_2024}. We used a Random Over Sampling algorithm in the Imbalanced Learn library \cite {imbalanced_learn}. Overampling was applied on inactive molecules, until an active/inactive ratio of 1:15 was obtained. The generated molecules were carefully checked and validated through the RDKit library \cite{Landrum_2010} to avoid introducing a classification bias due to the presence of chemically invalid molecules. 

The validation has been performed using the SanitazeMol \footnote{\url{https://www.rdkit.org/docs/index.html}} function of Rdkit as reported in the official cookbook. Specifically, each molecule was tested with the \textit{catch\_error} parameter set to \textbf{True} in order to receive output on any errors. In this way, every single Canonical Smiles generated was tested by evaluating the parameters listed in the table \ref{tab:sanitize_flags}.

\begin{table}[!ht]
\centering
\resizebox{.8\columnwidth}{!}{%
\begin{tabular}{l|p{10cm}}
\hline
\textbf{Flag} & \textbf{Description} \\
\hline
\texttt{SANITIZE\_CLEANUP} & Cleans the structure by removing invalid bonds and normalizing the molecule. \\
\hline
\texttt{SANITIZE\_PROPERTIES} & Checks and assigns atomic properties such as valences and formal charges. \\
\hline
\texttt{SANITIZE\_SYMMRINGS} & Identifies symmetric rings to prepare for aromaticity determination. \\
\hline
\texttt{SANITIZE\_KEKULIZE} & Converts aromatic bonds into a Kekulé representation (alternating single and double bonds). \\
\hline
\texttt{SANITIZE\_SETAROMATICITY} & Determines which atoms and bonds are aromatic. \\
\hline
\texttt{SANITIZE\_FINDRADICALS} & Identifies and assigns unpaired electrons (radicals) to atoms. \\
\hline
\texttt{SANITIZE\_ADJUSTHS} & Adjusts the number of implicit hydrogens based on the valence of the atoms. \\
\hline
\texttt{SANITIZE\_CLEANUPCHIRALITY} & Removes or corrects invalid or ambiguous stereochemistry. \\
\hline
\texttt{SANITIZE\_FINDSTEREO} & Identifies stereogenic centers and configures stereochemistry (e.g., R/S, E/Z). \\
\hline
\texttt{SANITIZE\_SETCONJUGATION} & Determines the conjugation of bonds (e.g., conjugated $\pi$ systems). \\
\hline
\texttt{SANITIZE\_SETHYBRIDIZATION} & Assigns hybridization to atoms (e.g., sp, sp$^2$, sp$^3$). \\
\hline
\texttt{SANITIZE\_ALL} & Performs \textbf{all} the checks listed above. \\
\hline
\end{tabular}}
\caption{RDKit Sanitization Flags and their Descriptions}
\label{tab:sanitize_flags}
\end{table}

\begin{figure}[!h]
    \centering
    \includegraphics[width=0.8\linewidth]{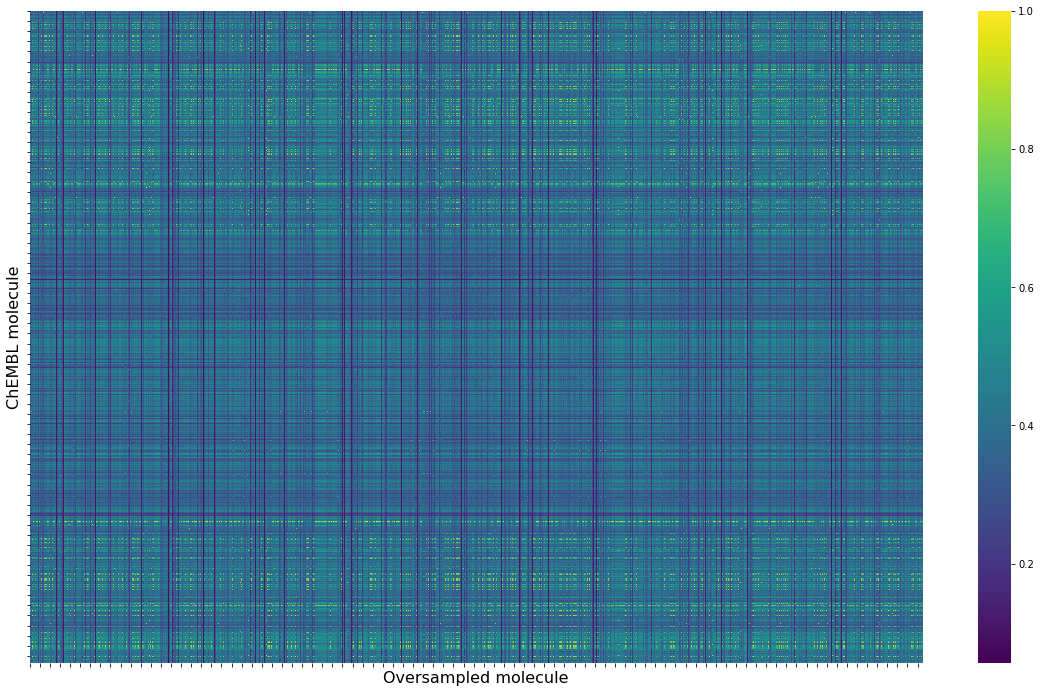}
    \caption{Heatmap of the similarity matrix obtained between the oversampled and ChEMBL molecules.}
    \label{fig:similarity-check}
\end{figure}

Once the validation phase had been completed, a global evaluation analysis of the new dataset was carried out. Specifically, a similarity study was performed between all the molecules generated with the oversampling procedure against the molecules downloaded from ChEMBL. This experiment allowed the construction of a similarity matrix, shown in figure \ref{fig:similarity-check}, where it is highlighted how the inclusion of these molecules leads to no bias in terms of similarity, and on the other hand brings out a remarkable structural uniformity.

Table \ref{tab:tab_dataset_2} reports the amount of inactive molecules in the ChEMBL\_over data set, arranged as ``Active'', ``I.C.'' (Inactive Compound) and ``I.O.'' (Inactive Oversampled).

\begin{table}[!h]
    \centering
    \resizebox{.3\columnwidth}{!}{%
    \begin{tabular}{l|c|c|c}
        \toprule
        Protein & Actives & I.C. & I.O. \\
        \midrule
        ACK1 & 816 & 282 & 12240 \\
        ALK & 2136 & 350 & 32040 \\
        CDK1 & 1310 & 769 & 19650 \\
        CDK2 & 2161 & 991 & 32385 \\
        CDK6 & 746 & 257 & 11190 \\
        CHK1 & 2475 & 684 & 22980 \\
        CK2A1 & 1145 & 174 & 18030 \\
        CLK2 & 786 & 263 & 99225 \\
        DYRK1A & 1549 & 565 & 12390 \\
        EGFR & 6947 & 2574 & 19755 \\
        ERK2 & 3735 & 721 & 37125 \\
        GSK3B & 3120 & 1252 & 17175 \\
        INSR & 1532 & 566 & 11790 \\
        IRAK4 & 2512 & 315 & 23235 \\
        ITK & 1202 & 183 & 104205 \\
        JAK2 & 6615 & 667 & 56010 \\
        JNK3 & 826 & 758 & 46800 \\
        MAPK2K1 & 1342 & 421 & 37680 \\
        MELK & 1317 & 290 & 20115 \\
        PDK1 & 1140 & 250 & 17100 \\
        \bottomrule
    \end{tabular}}
    \caption{Summary of active and inactive molecules in the ChEMBL\_over data set}
    \label{tab:tab_dataset_2}
    \end{table}

\subsection{Graph Convolutional Neural Network} \label{gcnn}
The neural architecture employed in this study consists of two main components: a set of Graph Convolutional (GC) blocks and the classifier, which is implemented using linear layers. 
Each of the 20 single-target networks follows the same general architecture. The GC blocks are responsible for processing the graph-structured input data, capturing relationships between nodes through convolutional operations. Specifically, each block consists of a dedicated GC layer that implements the \emph{1-dimensional Weisfeiler-Leman kernel} (1-WL), as presented in \cite{morris2019weisfeiler}, to enhance the network's ability in extracting meaningful structural features. 
The dense layers serve as the final classifier, mapping the extracted features to the output predictions.

It is well known \cite{Gimeno_Ojeda-Montes_Tomás-Hernández_Cereto-Massagué_Beltrán-Debón_Mulero_Pujadas_Garcia-Vallvé_2019} that the global efficacy of a machine learning model for Virtual Screening is measured through the ability of correctly prioritizing the \emph{True Positives}. 
Therefore, the main aim of the network is to provide a very high sensitivity across all the protein targets analyzed in this study. 
This focus on high sensitivity allows the network to deliver precise explanations of the molecular moieties involved in the target class assignment. To achieve this, we carefully optimized the network architecture, tailoring its structure to effectively balance complexity and performance.

The GC layers process the features in a U-Net-like fashion \cite{ronneberger2015u}, initially expanding and then reducing the number of channels layer by layer, and ultimately resulting in a final representation with 32 features. 
The architecture consists of seven GC blocks, with the number of channels in each GC layer following this sequence: 32, 64, 128, 256, 128, 64, and again 32. Layer normalization and the ReLU activation function are used after each block.

\begin{figure}
    \centering
    \captionsetup{justification=centering}
    \includegraphics[width=0.8\linewidth]{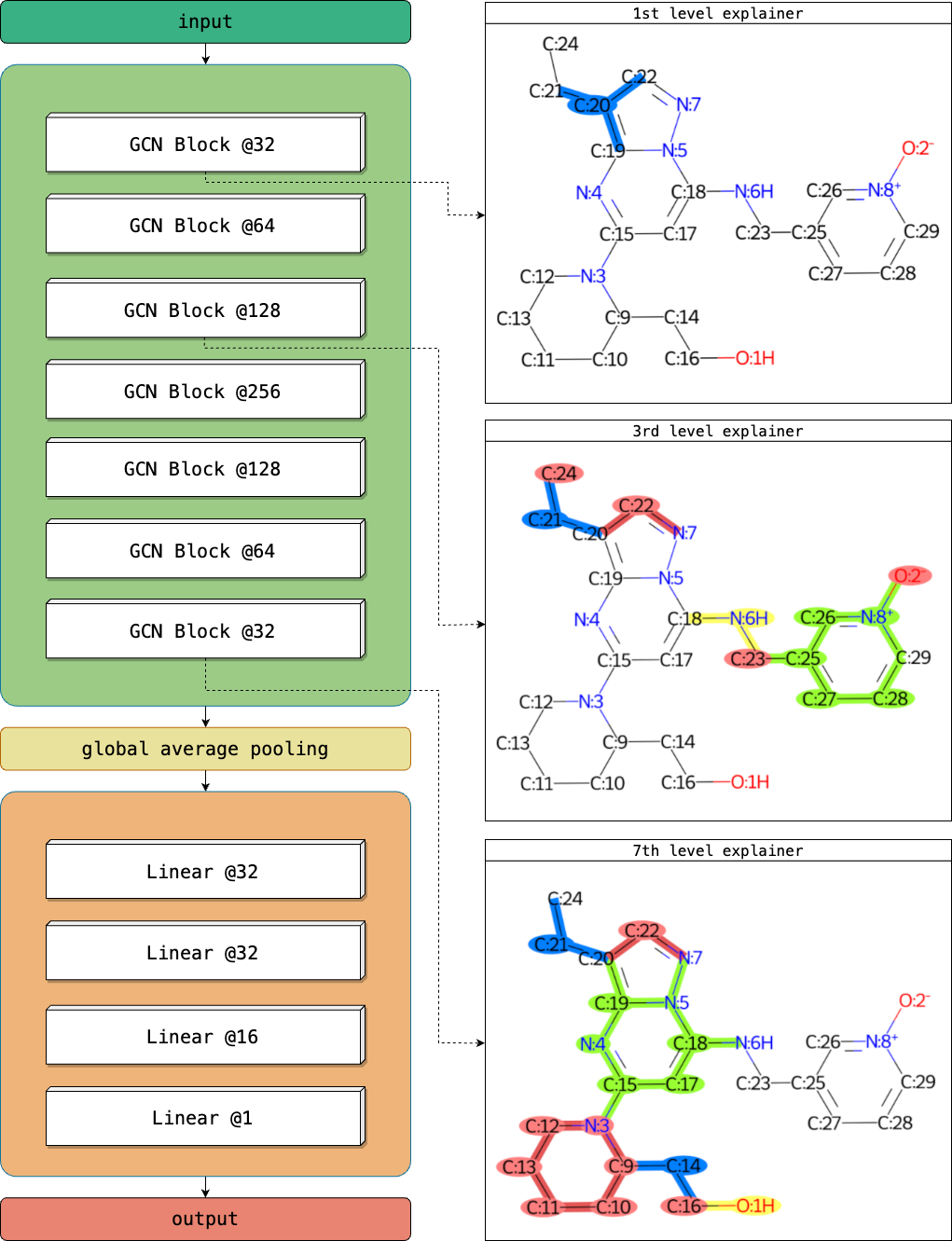}
    \vspace{2mm}
    \caption{The GCNN classifier architecture along with the HGE explainers}
    \label{fig:single-target-architecture}
\end{figure}

The convolutional layers described above use the message passing algorithm to stack data coming from the neighbors of each node in the graph. This aggregation process results in an updated set of node features placed along the channel dimension of the data sample. 
More formally, the algorithm uses a message function \(M_t\) and a vertex update function \(U_t\), which is usually implemented as a Multi-Layer Perceptron (MLP). Both \(M_t\) and \(U_t\) are learnable and differentiable functions to allow training in a neural network. 
Given the time-step \(t\), the node \(v\) collecting features, the edge \(e_{vw}\) connecting node \(v\) and any node \(w\) in the set \(N(v)\) of the neighbors of \(v\) in the graph \(G\), and the hidden states \(h_v^t\) and \(h_w^t\), the new message \(m_v\) at time-step \(t+1\) is calculated as:

\begin{equation}
m_v^{t+1} = \sum_{w\in N(v))} M_t( h_v^t, h_w^t, e_{vw})
\label{eq:message_function}
\end{equation}

The new hidden state \(h_v^{t+1}\) is then computed from \(h_v^t\) and the new message \(m_v^{t+1}\) using the vertex update function:

\begin{equation}
    h_v^{t+1} = U_t(h_v^t, m_v^{t+1})
\label{eq:vertex_update}
\end{equation}

\vspace{5pt}
The augmented node feature vector coming from the GC operations flows into a Global Average Pooling (GAP) layer. This step is also known as the \textit{readout phase}, and it represents a core operation when using the message passing algorithm. GAP condenses the information about all the nodes into a one-dimensional vector that averages all the features of the whole graph into a compressed and meaningful vector representation.

The second part of the network is a neural classifier consisting of four fully connected layers whose channels are, respectively, 32, 32, 16, and finally 1 that is fed to a sigmoid to estimate the probability for the input to be active on the target. ReLU activation functions and dropout layers are also added for each fully connected layer to add non-linearity correlation to the input data. The entire architecture of the single target network is reported in Figure~\ref{fig:single-target-architecture} along with an example of the output of the explainers used in HGE.

A separate GCNN binary classifier was trained to assess compounds activity for each target. Weighted Binary Cross-Entropy was used with a higher weight value for the positives given the strong imbalance between active and inactive classes on targets. A grid search was performed to select the hyperparameters properly. The number of GC filters was searched in the set $\{1024, 512, 256, 128, 64, 32, 16\}$, while learning rates were in the range $[10^{-5}, 10^{-1}]$. The batch size ranged from $[8, 64]$. The best model was stored after each epoch, and the early stopping callback was used to determine the ideal number of training epochs. 
The training process was conducted on an NVIDIA GeForce RTX 3090 with 10,496 CUDA cores, handling the extensive computations required for the grid search evaluation to determine the optimal hyperparameters.

\subsection{Explainability analysis}\label{xai_method}
As mentioned previously, the main contribution of this work is our Hierarchical Grad-CAM graph Explainer devised to gain information about the most relevant molecular moieties involved in protein-ligand binding prediction; different algorithms were tried in this respect. 
All selected XAI approaches belong to the so-called ``instance-level methods'', that is, the ones that aim to identify the input features that contribute mainly to the prediction. At first, we applied GNNExplainer \cite{ying2019gnnexplainer}, a ``perturbation-based method'' that tracks the change in the output for minor variations of the same input data to determine which features are the most important for prediction.
GNNExplainer provides information at a very fine-grained level, focusing on individual features. However, our goal was to obtain explanations at a higher level, specifically at the atom (i.e. node) level. As a consequence, we moved towards the class of ``gradient and feature-based techniques''. An implementation from the work by Pope et al. \cite{Pope2019ExplainabilityMF} was utilized for this purpose.
In this work, the authors implemented some graph-based counterparts of three well-established explainability methods, which had originally been designed for convolutional neural networks.

We chose Grad-CAM as the best performing technique for labeling pointwise contributions to positive predictions, as it provides valuable information on the most relevant features driving the model classification, particularly highlighting the key pharmacophoric elements within molecular graphs, as already mentioned in Section \ref{sec1}. This ability to visualize the model’s decision-making process, in addition to its suitability in the original image domain, where pixels are the data units and information flows between layers via the convolution operator, made it an ideal choice.
In a GC layer, feature maps are built for each node of the graph by message passing (eq. \ref{eq:message_function}). In this way, the features of each node encode the information flow from the node neighborhood, and the deeper the layer, the more its features collect information from many message passing steps computed in the previous layers (eq. \ref{eq:vertex_update}). As a consequence, a feature map at a certain depth $d$ collects information from the $d-$neighbors of a node.

Grad-CAM aims at visualizing the most active region of the input when a particular class is predicted. This is done by examining the gradients of the target class score with respect to each feature map in the network layer under investigation. The weight $\alpha _k ^ {l,c}$, for class $c$ related to the $k$-th node feature at layer $l$ is computed as:
\begin{equation} 
\alpha _k ^ {l,c} = \frac{1}{N} \sum_{n=1}^{N} \frac{\partial y^c}{\partial F_{k,n}^l}
\label{eq:alphas}
\end{equation} 
$F_{k, n}^l$ is the $k$-th feature value for the single node $n$ in layer $l$, while $y^c$ is the score for class $c$ obtained from the last network layer, before the output unit.  Obviously, the partial derivative term is computed using the chain rule when $l$ is not the last layer. Grad-CAM values for the entire node $n$ are summed over its features, and passed to the ReLU activation function to filter out negative contributions to the prediction of class $c$.

\begin{equation} 
L_{Grad-CAM} [l, n] = \operatorname{ReLU}\left(\sum_{k} \alpha _k ^ {l,c} F_{k, n}^l\right)
\label{eq:Grad-CAM_values}
\end{equation} 
\vspace{2mm}

In our implementation, we set $c = 1$, that is, we have a single class score, as the explainer was developed for a binary classifier. Moreover, the ReLU activation function has been removed to capture not only positive contributions to the prediction, but also negative ones. Our choice is motivated by the intrinsic nature of the classification task to be explained. 
In the image domain, classification tasks can involve assigning each pixel to a specific class, as in semantic segmentation, or labeling the entire image based on the presence of class-discriminative features learned by the network. In both cases, classification can be interpreted as a pattern recognition task. Consequently, the explainer is designed to highlight ``regions'' that contribute to the prediction while disregarding irrelevant areas, which is the intended purpose of the ReLU activation function.

In this context, bioactivity prediction is a classification task in which molecules are labeled positive or negative if they ``globally'' trigger a particular biological event or not. Many factors are involved in bioactivity: the presence of suitable pharmacophoric moieties, molecular chirality, solubility, etc. All of them can be captured by training the GCNN on a three-dimensional molecular representation with rich information at the nodes, but cannot be directly related to just ``positive'' or ``negative'' Grad-CAM values at each node. As a consequence, we removed ReLU for our explainability task. Moreover, pharmacophoric information is strictly related to the atom connectivity within the molecular structure. 

In view of this last consideration, we evaluated Grad-CAM multiple times at different depths in the convolutional segment of the network. We used the classical Grad-CAM arrangement after the last layer of the network, which is the seventh. This explainer aims to highlight relevant features globally and is unable to provide information about subtle chemical and pharmacophoric moieties like H-bond acceptors or donors. To reach this aim, we placed also a Grad-CAM explainer at the very first GC layer to consider the atoms' impact in a 1-neighborhood that is at a bond level. Another explainer was placed after the third layer to catch information related to 3-neighborhoods. At this level, as we analyze the 3-neighborhood of each atom, the action of an entire substructure like aromatic ring can be explained. 

The whole explanation map has been built as follows. At first, the three Grad-CAM value distributions have been summed for each node: 

\begin{equation} 
L_{Grad-CAM} [n] = L_{Grad-CAM} [1, n] + L_{Grad-CAM} [3, n] + L_{Grad-CAM} [7, n]
\label{eq:global_Grad-CAM_values}
\end{equation} 
The resulting $L_{Grad-CAM}$ distribution has been min-max scaled, and only contributions higher than $0.7$ have been regarded as relevant. 
We called the final output ``merged output''. It maps all of the key chemical features collected from the various explainer layers, combining them to create a global view of each of these levels' contributions. This approach will determine the whole chemical interacting feature assignment.
The GCN architectures are in fact able to recognize molecular graphs and classify the molecules based on their nodes and connections. However, atomic typing and identification of the most important chemical groups in determining the assignment of a molecule to a specific target are lacking. 
In order to account this issue, we employed HGE explanations to contextualise the information obtained. To achieve this, it was necessary to harness the power of the software package RDKit\footnote{\url{https://www.rdkit.org/}}. Actually, the software has the capability to identify pharmacophoric regions for each of the molecular sub-structures found in the molecular structure. 
Once the moieties are labeled, they are cross-referenced with the results obtained from HGE to select only those that contributed to the bioactivity classification. This procedure is crucial for each layer of the explainer and allows us to chemically translate the information obtained from the various layers of the GCN.
Looking at Figure \ref{fig:single-target-architecture}, H-bond acceptor or donor groups are highlighted in yellow, hydrophobic groups are colored in blue, and the aromatic groups are green. Lastly, moieties without an explicit pharmacophoric label where $L_{Grad-CAM} > 0.7$ are labeled as ``relevant'', and are colored in red. In Section \ref{exp-results} a detailed analysis is reported about the structures highlighted by HGE.

\section{Results}\label{sec3}
Experiments were arranged in three stages. At first, a GCNN classifier was trained for each target, to allow HGE computation. Furthermore, the explainability analysis was performed on 143 known kinase inhibitors from DrugBank\footnote{https://go.drugbank.com/)}, and about 80\% of them were correctly assigned to the annotated target.
Finally, our GCNN architectures were compared with Chemprop and Attentive FP \cite{Zheng_2024, Chen_Wang_2024, Zhang_Ren_2025}, two of the most widely used frameworks for training message-passing neural networks aimed at predicting molecular properties.

\subsection{Metrics}\label{metrics}
To test the predictive capabilities of the proposed architectures, we used two families of metrics. At first we reported the classical Machine Learning scores used for classification.
The \textit{Balanced Accuracy} assesses the network's ability to make correct predictions despite the class imbalance, while the \textit{Sensitivity} (also named \textit{Recall}) focuses solely on the network's performance in correctly classifying active molecules.
The \textit{F1-score} quantifies how well the model can correctly identify positive instances (\textit{Recall}) while avoiding false positives (\textit{Precision}). On the other hand, the \textit{Area Under the Curve} (AUC) provides a global measure of the model's ability to discriminate between positive and negative instances, and it is computed from the Receiver Operating Characteristics (ROC) curve that plots the Sensitivity vs. the False Positive Rate varying the threshold to discriminate positives and negatives.

Prioritizing the best active ligands is the true outcome of a VS task. For this purpose, the \textit{True Positive vs Positives} ratio (TP/P) and the \textit{Enrichment Factor} (EF) are used. TP/P is employed to assess the correct predictions count relative to the total positives in a given class, the EF metric measures how many times the tested model outperforms a purely random process in predicting a given class. Both measures are computed for a given topmost percentage of the test set. We used, respectively, 1\%, 2\%, 5\%, and 10\% of the whole test set, according to the chemical literature.

\subsection{Results on EMBER data set}\label{dataset_1_results}
The results on the EMBER data set for each single-target classifier are shown in Table \ref{tab:single_target_dataset_1} and Table \ref{tab:tpp_ef_1_target}. Table \ref{tab:single_target_dataset_1} shows the results achieved by our architecture for each target, and the last three columns show the sensitivity comparison with Chemprop and Attentive FP. In addition, a comparison with Chemprop and Attentive FP is reported in Table \ref{tab:comparison_metrics_dataset1} where all metrics averaged on the 20 targets are reported for both approaches. 

As it can be seen, F1-Score values show that the network is able to overcome the data set imbalance while simultaneously reducing the number of false negatives and false positives. The use of a properly balanced loss and a soft F-score worked fine for the classification task. Particularly noticeable is the Sensitivity value that is higher than that of Chemprop for all the targets.

\begin{table}[!h]
    \caption{Performance metrics of our GCNN architecture on EMBER data set.}
    \centering
    \begin{tabular}{cccc||ccc}
    \toprule
\textbf{Protein} & \textbf{Bal. Accuracy} & \textbf{F1-Score} & \textbf{AUC} & \textbf{G-Sensitivity}&\textbf{C-Sensitivity}&\textbf{A-Sensitivity} \\ \midrule
ACK              & 0.891                     & 0.979             & 0.983                             & \textbf{0.811}&0.657&0.0822                  \\
ALK              & 0.929                     & 0.972             & 0.987                             & \textbf{0.893}&0.742&0.2532                  \\
CDK1             & 0.921                     & 0.956             & 0.969                             & \textbf{0.907}&0.650&0.2650                  \\
CDK2             & 0.932                     & 0.940             & 0.971                             & \textbf{0.950}&0.721&0.4099                  \\
CDK6             & 0.916                     & 0.989             & 0.988                             & \textbf{0.846}&0.772&0.3158                  \\
CHK1             & 0.923                     & 0.954             & 0.977                             & \textbf{0.912}&0.604&0.6443                  \\
CK2A1            & 0.942                     & 0.979             & 0.989                             & \textbf{0.911}&0.705&0.3786                  \\
CLK2             & 0.938                     & 0.929             & 0.989                             & \textbf{0.965}&0.881&0.0794                  \\
DYRK1A           & 0.913                     & 0.983             & 0.986                             & \textbf{0.847}&0.730&0.2376                  \\
EGFR             & 0.930                     & 0.982             & 0.990                             & \textbf{0.881}&0.749&0.8859                  \\
ERK2             & 0.927                     & 0.941             & 0.977                             & \textbf{0.935}&0.782&0.7416                  \\
GSK3B            & 0.903                     & 0.970             & 0.978                             & \textbf{0.847}&0.615&0.6639                  \\
INSR             & 0.893                     & 0.974             & 0.977                             & \textbf{0.823}&0.430&0.2279                  \\
IRAK4            & 0.930                     & 0.953             & 0.979                             & \textbf{0.931}&0.619&0.4862                  \\
ITK              & 0.928                     & 0.921             & 0.981                             & \textbf{0.958}&0.814&0.1625                  \\
JAK2             & 0.945                     & 0.945             & 0.984                             & \textbf{0.960}&0.786&0.9000                  \\
JNK              & 0.918                     & 0.923             & 0.971                             & \textbf{0.946}&0.691&0.0154                  \\
MAP2K1           & 0.944                     & 0.969             & 0.991                             & \textbf{0.926}&0.807&0.2381                  \\
MELK             & 0.916                     & 0.969             & 0.982                             & \textbf{0.874}&0.635&0.5229                  \\
PDK1             & 0.913                     & 0.973             & 0.986                             & \textbf{0.860}&0.693&0.3818                  \\ 
    \midrule
     & & & \textbf{mean}&\textbf{0.899}&0.704&0.3946\\
    \bottomrule
    \end{tabular}
    \footnotesize *\textbf{G-Sensitivity} = Our GCN Sensitivity; \textbf{C-Sensitivity} = Chemprop Sensitivity; \textbf{A-Sensitiviy} = Attentive FP Sensitivity. 
    \label{tab:single_target_dataset_1}
\end{table}

\begin{table}[!ht]
\resizebox{\columnwidth}{!}{%
\begin{tabular}{@{}lcccccccc@{}}
\toprule
\textbf{Protein} & \textbf{TP/P 1\%} & \textbf{TP/P 2\%} & \textbf{TP/P 5\%} & \textbf{TP/P 10\%} & \textbf{EF 1\%} & \textbf{EF 2\%} & \textbf{EF 5\%} & \textbf{EF 10\%} \\ 
\midrule
ACK              & 72/106            & 78/106            & 94/106            & 100/106            & 67              & 36              & 17              & 9                \\
ALK              & 123/254           & 186/254           & 227/254           & 242/254            & 48              & 36              & 17              & 9                \\
CDK1             & 80/205            & 117/205           & 177/205           & 190/205            & 39              & 28              & 17              & 9                \\
CDK2             & 86/303            & 141/303           & 240/303           & 287/303            & 28              & 23              & 15              & 9                \\
CDK6             & 82/104            & 89/104            & 94/104            & 99/104             & 78              & 42              & 18              & 9                \\
INSR             & 89/217            & 123/217           & 180/217           & 204/217            & 41              & 28              & 16              & 9                \\
ITK              & 98/158            & 127/158           & 145/158           & 154/158            & 62              & 40              & 18              & 9                \\
JAK2             & 134/832           & 268/832           & 651/832           & 774/832            & 16              & 16              & 15              & 9                \\
JNK3             & 79/105            & 83/105            & 93/105            & 102/105            & 75              & 39              & 17              & 9                \\
MELK             & 125/185           & 158/185           & 171/185           & 179/185            & 67              & 42              & 18              & 9                \\
CHK1             & 126/343           & 189/343           & 269/343           & 319/343            & 36              & 27              & 15              & 9                \\
CK2A1            & 93/151            & 115/151           & 128/151           & 141/151            & 61              & 38              & 16              & 9                \\
CLK2             & 52/102            & 63/102            & 85/102            & 96/102             & 50              & 30              & 16              & 9                \\
DYRK1A           & 72/174            & 106/174           & 144/174           & 167/174            & 41              & 30              & 16              & 9                \\
EGFR             & 129/702           & 262/702           & 520/702           & 641/702            & 18              & 18              & 14              & 9                \\
ERK2             & 130/525           & 260/525           & 444/525           & 503/525            & 24              & 24              & 16              & 9                \\
GSK3             & 102/393           & 174/393           & 293/393           & 358/393            & 25              & 22              & 14              & 9                \\
IRAK4            & 134/339           & 249/339           & 311/339           & 327/339            & 39              & 36              & 18              & 9                \\
MAP2K1           & 106/191           & 133/191           & 166/191           & 181/191            & 55              & 34              & 17              & 9                \\
PDK1             & 108/187           & 132/187           & 164/187           & 183/187            & 57              & 35              & 17              & 9                \\ \bottomrule
\end{tabular}%
}
\caption{TP/P and EF results for the EMBER data set}
\label{tab:tpp_ef_1_target}
\end{table}

\begin{table}[!ht]
\centering
\begin{tabular}{c c c c c}
\hline
    \textbf{Model} & \textbf{Bal. Accuracy} & \textbf{Sensitivity} & \textbf{F1-Score} & \textbf{AUC}\\
\hline
    Chemprop & 0.851 & 0.704 & 0.756 & 0.851\\
    Attentive FP & 0.681 & 0.427&	0.935&	0.896\\
    Our GCNN & \textbf{0.923} & \textbf{0.899} & \textbf{0.960} & \textbf{0.982} \\ 
\hline
\end{tabular} %
\caption{Average evaluation metrics for Chemprop, Attentive FP and our GCNN on EMBER data set}\label{tab:comparison_metrics_dataset1}
\end{table}

As can be seen in table \ref{tab:single_target_dataset_1}, our architecture outperforms in terms of sensitivity the two architectures used as a comparison on all 20 protein targets for the EMBER dataset. This result highlights the excellent capabilities of our GCN in the correct classification of small molecules and as can be seen from the table \ref{tab:tpp_ef_1_target} it also has an very good ability to prioritise True Positives in terms of TP/P and EF.
Finally, as shown in table \ref{tab:comparison_metrics_dataset1} our GCN, overall performs better in terms of F1-score and AUC on the average of each target.

\subsection{Results on ChEMBL\_over data set}
We carried out the same experiments on the ChEMBL\_over data set as on the EMBER one. The results obtained from the 20 GCNNs trained purposely are reported in table \ref{tab:res_dataset_2}, table \ref{tab:tpp_ef_2_target}, and table \ref{tab:comparison_metrics_dataset2}.

\begin{table}[!ht]
    \caption{Performance metrics of our GCNN architecture on ChEMBL\_over data set.}
    \centering
    \begin{tabular}{cccc||ccc}
    \toprule
    \textbf{Protein}&\textbf{Bal. accuracy}&\textbf{AUC}&\textbf{F1-score}&\textbf{G-Sensitivity*}&\textbf{C-Sensitivity*}&\textbf{A-Sensitivity}\\
    \midrule
    ACK1&0.972&0.992&0.954&\textbf{0.947}&0.866&0.921\\
    ALK&0.979&0.996&0.978&\textbf{0.957}&0.944&0.934\\
    CDK1&0.949&0.995&0.933&0.901&\textbf{0.939}&0.832\\
    CDK2&0.953&0.985&0.862&\textbf{0.929}&0.907&0.862\\
    CDK6&0.952&0.951&0.901&\textbf{0.914}&0.893&0.814\\
    CHK1&0.969&0.991&0.965&\textbf{0.939}&0.907&0.905\\
    CK2A1&0.986&1.000&0.986&\textbf{0.972}&0.896&0.926\\
    CLK2&0.956&0.991&0.954&\textbf{0.911}&0.899&0.861\\
    DYRK1A&0.943&0.974&0.871&\textbf{0.903}&0.723&0.883\\
    EGFR&0.914&0.961&0.715&0.887&\textbf{0.895}&0.870\\
    ERK2&0.954&0.979&0.803&\textbf{0.950}&0.893&0.922\\
    GSK3B&0.958&0.984&0.876&\textbf{0.937}&0.869&0.906\\
    INSR&0.981&0.990&0.981&\textbf{0.962}&0.928&0.920\\
    IRAK4&0.993&0.999&0.993&\textbf{0.986}&0.952&0.970\\
    ITK&0.997&1.000&0.997&\textbf{0.995}&0.950&0.968\\
    JAK2&0.993&0.998&0.980&\textbf{0.989}&0.970&0.975\\
    JNK3&0.942&0.979&0.707&\textbf{0.959}&0.819&0.803\\
    MAPK2K1&0.964&0.989&0.963&\textbf{0.928}&0.903&0.871\\
    MELK&0.992&0.999&0.992&\textbf{0.984}&0.833&0.972\\
    PDK1&0.988&1.000&0.988&\textbf{0.977}&0.947&0.954\\
    \midrule
     & & & \textbf{mean}&\textbf{0.946}&0.897&0.904\\
    \bottomrule
    \end{tabular}
    \footnotesize *\textbf{G-Sensitivity} = Our GCN Sensitivity; \textbf{C-Sensitivity} = Chemprop Sensitivity; \textbf{A-Sensitivity} = Attention FP Sensitivity.
    \label{tab:res_dataset_2}
\end{table}

\begin{table}[!h]
    \resizebox{\columnwidth}{!}{%
    \begin{tabular}{@{}lcccccccc@{}}
    \toprule
    \textbf{Protein} & \textbf{TP/P 1\%} & \textbf{TP/P 2\%} & \textbf{TP/P 5\%} & \textbf{TP/P 10\%} & \textbf{EF 1\%} & \textbf{EF 2\%} & \textbf{EF 5\%} & \textbf{EF 10\%} \\ 
    \midrule
ACK1&9/76&17/76&44/76&73/76&12&11&12&10\\
ALK&23/211&46/211&115/211&202/211&11&11&11&10\\
CDK1&14/131&29/131&72/131&125/131&11&11&11&10\\
CDK2&25/225&49/225&123/225&208/225&11&11&11&9\\
CDK6&8/70&16/70&39/70&64/70&11&11&11&9\\
CHK1&28/263&57/263&142/263&248/263&11&11&11&9\\
CK2A1&12/108&25/108&62/108&108/108&11&12&11&10\\
CLK2&9/79&18/79&44/79&75/79&11&11&11&10\\
DYRK1A&17/154&34/154&84/154&140/154&11&11&11&9\\
EGFR&71/655&143/655&347/655&539/655&11&11&11&8\\
ERK2&40/361&79/361&198/361&332/361&11&11&11&9\\
GSK3B&54/491&109/491&271/491&460/491&11&11&11&9\\
INSR&28/262&57/262&142/262&257/262&11&11&11&10\\
IRAK4&41/368&81/368&204/368&365/368&11&11&11&10\\
ITK&21/185&41/185&104/185&185/185&11&11&11&10\\
JAK2&112/1040&225/1040&562/1040&1031/1040&11&11&11&10\\
JNK3&13/122&27/122&66/122&110/122&11&11&11&9\\
MAPK2K1&22/194&45/194&111/194&189/194&11&12&11&10\\
MELK&38/364&76/364&190/364&361/364&10&10&10&10\\
PDK1&20/173&39/173&98/173&173/173&12&11&11&10\\
\bottomrule
    \end{tabular}}
    \caption{TP/P and EF results for the ChEMBL\_over data set}
    \label{tab:tpp_ef_2_target}
\end{table}

\begin{table}
\centering
\begin{tabular}{c c c c c}
\hline
    \textbf{Model} & \textbf{Bal. Accuracy} & \textbf{Sensitivity} & \textbf{F1-Score} & \textbf{AUC}\\
\hline
    Chemprop & 0.948 & 0.897 & \textbf{0.945} & 0.982\\
    Attentive FP& 0.9450&	0.904&	0.887&	0.975\\
    Our GCNN & \textbf{0.967} & \textbf{0.946} & 0.920 & \textbf{0.988} \\ 
\hline
\end{tabular} %
\caption{Average evaluation metrics for Chemprop, Attentive FP and our GCNN on ChEMBL\_over data set.}\label{tab:comparison_metrics_dataset2}
\end{table}

The GCNNs' performance on the ChEMLB\_over data set is consistent with that obtained on the EMBER one.  Our architectures outperform Chemprop on 18 out of 20 targets in terms of sensitivity, but the difference between the two is quite minor.  The TP/P and EF ratings are strong across all objectives.  Finally, Chemprop's average F1-score is higher than our architectures.  This is not a constraint because our designs were trained to achieve high Sensitivity, as evidenced by the average of such a statistic across all targets. As a result, our networks outperform Chemprop at prioritising True Positive.  This result may suggest a vulnerability in the Chemprop net, which appears more prone than our GCNNs to be misled in determining True Positives by the presence of over-sampled chemicals. Despite the different results on the two targets (CDK1 and EGFR) compared to Chemprop, in relation to the results obtained by attentive FP, our architecture discriminates better on all targets. This further provides additional proof of the robustness of the proposed GCN against two of the best-performing deep neural architectures in terms of classification.

\subsection{Explainability results}\label{exp-results}

To evaluate the explainability power of HGE, we selected a test set made by 143 known drugs from DrugBank \footnote{https://go.drugbank.com/)} for the 20 kinases considered in this work to compare the predictions with experimental and clinical results. For this purpose, only drugs flagged as \textit{approved} and \textit{inhibitor} on DrugBank were chosen. For each drug chosen, correct target assignation was evaluated and the relevant chemical moieties were collected and analysed. For each chemotype and for each target, the consistency of key chemical features identification was assessed. 
Thus, 116 drugs out of 143 were correctly assigned to the target known in the literature. 
In order to further stress the discriminative capacity of the proposed approach, 13 drugs with bioactivity on multiple targets have been identified. The results obtained from the evaluation phase of the results are shown in table \ref{tab:preds_multitarget}. Finally, to further stress the capabilities of the proposed explainer, 13 drugs with multi-target bioactivity were tested to see if the explainer's predictions were kept consistent among different targets annotated. The results is shown in table \ref{tab:preds_multitarget}.

\begin{table}[!h]
    \centering
    \begin{tabular}{c|c|c}
    \toprule
    \textbf{Target}&\textbf{\# of Drugs}&*C.P.D.\\
    \midrule
    ACK&1&0\\
    CDK2&5&4\\
    CDK6&3&3\\
    CHK1&6&4\\
    C2KA1&4&4\\
    CLK2&1&1\\
    DYRK1A&1&1\\
    ERK2&5&1\\
    GSK3b&7&4\\
    INSR&1&1\\
    IRAK4&1&0\\
    ITK&2&2\\
    JAK2&2&2\\
    JNK3&1&1\\
    MAPK2K1&2&1\\
    MELK&1&0\\
    PDK1&1&1\\
    \bottomrule
    \end{tabular}
    \footnotesize *C.P.D. = Correctly Predicted Drugs
    \caption{Classification and explainability predictions on the test drugs}
    \label{tab:preds_multitarget}
\end{table}

\begin{figure}
    \centering
    \includegraphics[width=0.8\linewidth]{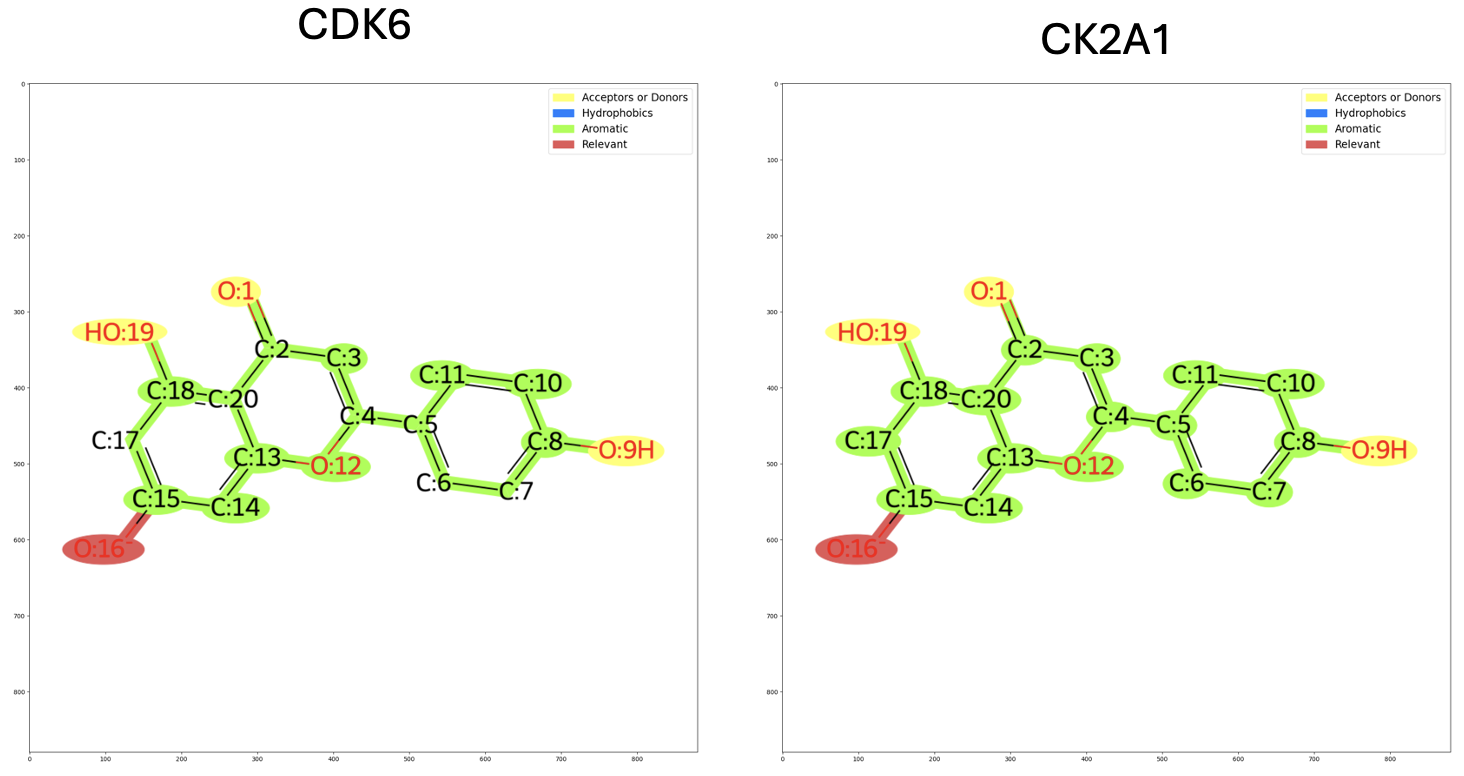}
    \caption{Predictions obtained with HGE for Apigenin on CDK6 and CK2A1. H-bond acceptor or donor groups are highlighted in yellow, hydrophobic groups are colored in blue, and the aromatic groups are green.}
    \label{fig:multitarget-apigenin}
\end{figure}

As shown in the table \ref{tab:preds_multitarget}, only 30/44 (0.6818) drugs with multi-target activity are correctly classified, but this result can be interpreted positively. In fact, the same molecules were tested with HGE, demonstrating the robustness in correctly classifying molecular scaffolds. Instead, the misclassification on the 14 multi-target drug molecules can be traced back to different reasons. From the analysis carried out, certainly, as shown by the heatmap in Fig 1. The molecules with which the model is trained do not present great chemical variability of the scaffolds, as is to be expected for kinase inhibitors with a similar mechanism of action and which therefore have quite similar pharmacophoric portions. For multi-target drugs, this structural similarity further complicates the possibility of recognizing motifs that can perfectly discriminate the molecules for a target. A second aspect that must certainly be underlined is that for some targets, training datasets contain fewer known molecules (see table 2), therefore the model has fewer examples on which to be trained and especially in the case of multi-target drugs shows lower performances. All the drugs have been tested with HGE, in order to assess whether the core pharmacophore could be predicted correctly, and these results are reported in supplementary material 3. 
Some examples are shown in figure \ref{fig:multitarget-apigenin}, where the prediction of the molecular moieties of Apigenin, a drug that is bioactive on CDK6 and CK2A1, remains congruous, despite the bioactivity is calculated on two different targets. This result demonstrates that HGE is consistent and robust when discriminating between compounds even when the neural network is trained on different protein targets.

\begin{figure}[!ht]
    \centering
    \includegraphics[width=0.8\linewidth]{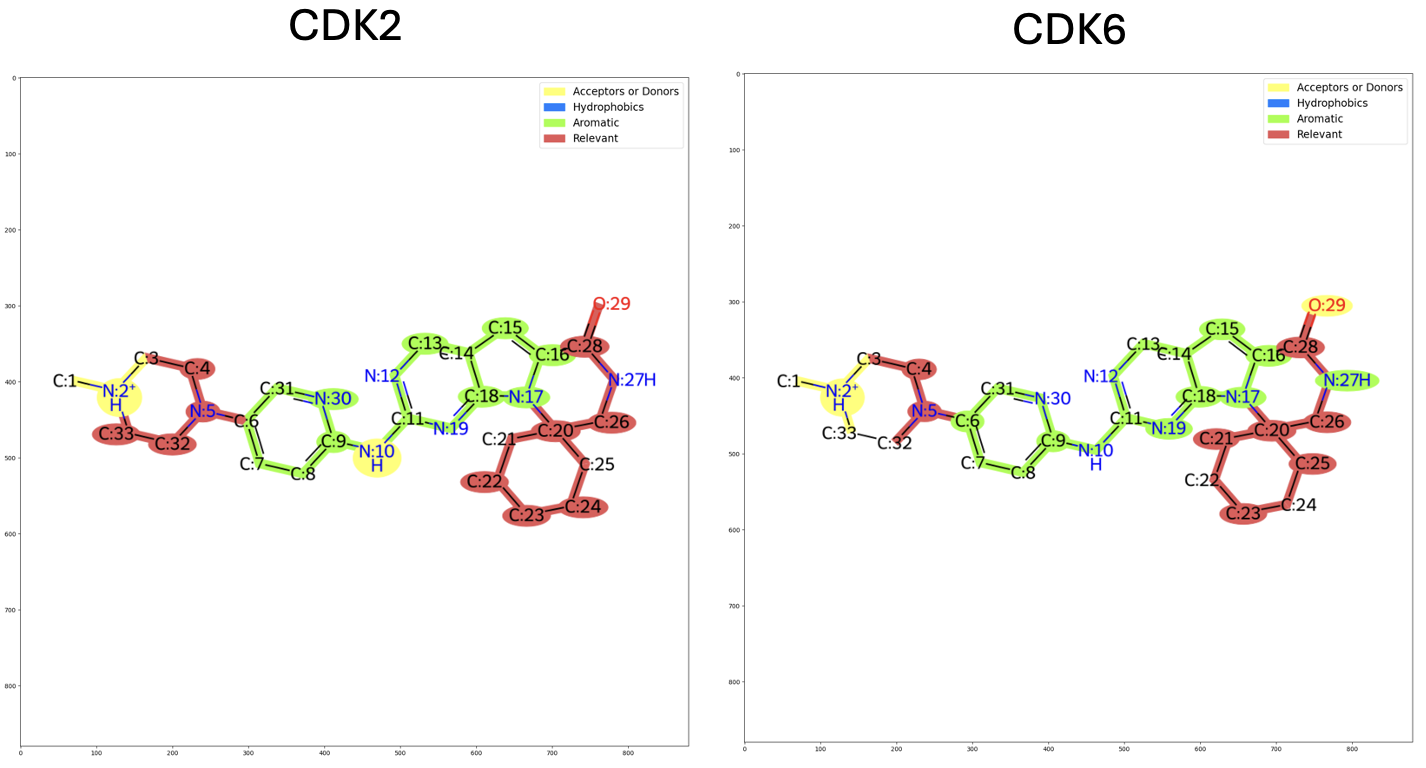}
    \caption{Predictions obtained with HGE for Trilaciclib on CDK2 and CDK6. H-bond acceptor or donor groups are highlighted in yellow, hydrophobic groups are colored in blue, and the aromatic groups are green.}
    \label{fig:multitarget-trilaciclib}
\end{figure}

Figure \ref{fig:multitarget-trilaciclib} shows the prediction of HGE for Trilaciclib as another example of this consistent behavior. This drug is active on both CDK2 and CDK6 whose binding sites are very similar. This evidence remains congruous for all the other 11 drugs (see supplementary material S3), demonstrating how HGE manages to maintain its predictive capacity of molecular moieties.

\begin{figure}[!h]
    \centering
    \includegraphics[width=\linewidth]{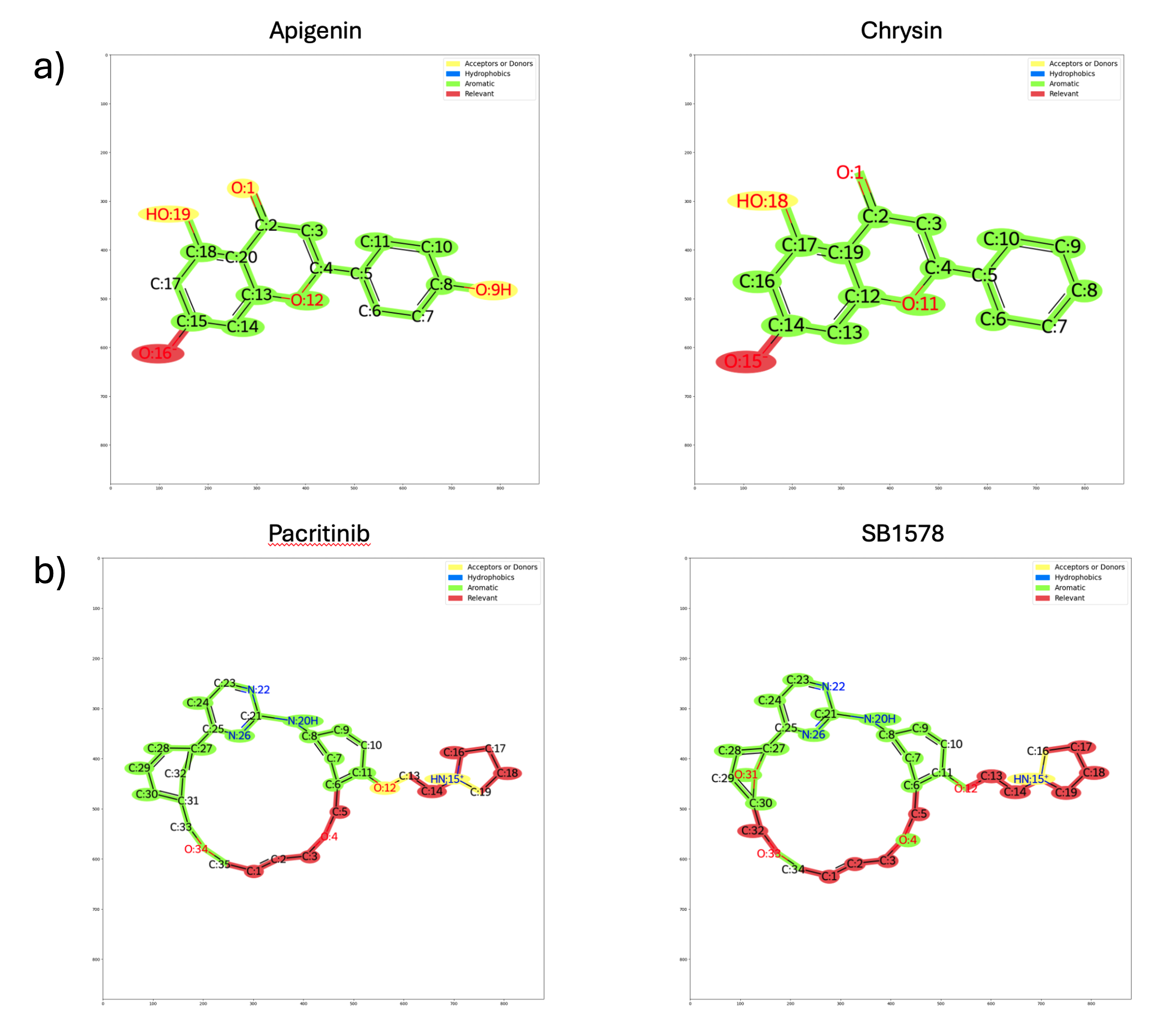}
    \caption{Chemical features identified by HGE in a) Apigenin and Chrysin when assigned to CDK6 protein and b) Chemical features identified by HGE in Pacritinib and SB1578 when assigned to JAK2 protein. H-bond acceptor or donor groups are highlighted in yellow, hydrophobic groups are colored in blue, and the aromatic groups are green.}
    \label{fig:xai_CDK6_JAK2}
\end{figure}

Considering the explainer's ability to correctly identify the moieties for drugs with multi-target activity, it was been decided to further test the robustness of HGE. For this phase, different drugs that are active on the same target have been identified  whose molecular scaffold was chemically similar. Some examples are shown in figure  \ref{fig:xai_CDK6_JAK2} which highlights the predicted moieties in Apigenin and Chrysin (both active on CDK6) and in Pacritinib and SB1578 (both active on JAK2).

\begin{figure}[!ht]
    \centering
    \includegraphics[width=0.8\linewidth]{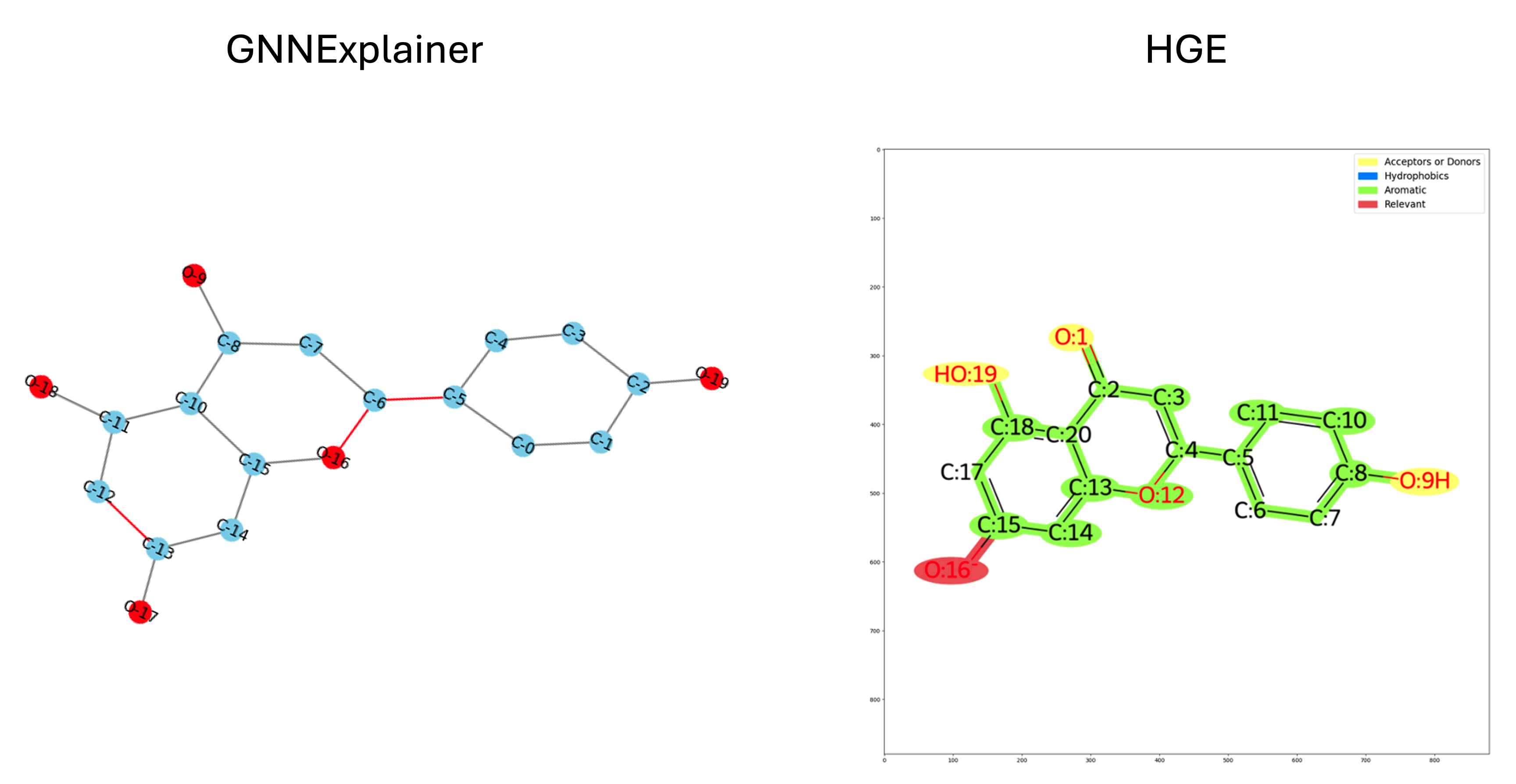}
    \caption{Results obtained with GNNExplainer compared with HGE on the Apigenin molecule.}
    \label{fig:GNNExplainerVsHGE}
\end{figure}

In order to further validate the abilities of our explainer, a comparison has been carried out with the GNNExplainer \cite{ying2019gnnexplainer}. This explainer is one of the most widely used for graph explaining procedures. The experiments performed involved all drugs selected for testing using the best neural architecture for each target. GNNExplainer allows, through a training procedure, to obtain the \textit{node-feat-mask} that describes the contribution of the features present in each node and an \textit{edge-mask} that instead highlights the arcs relevant for obtaining the task. Using the \textit{visualise\_graph} method, this explainer proposes a visualisation of the result applied directly to the generalist graph. Considering the pharmaceutical domain such as the one being studied, this visualisation does not make it easy to interpret the outcome. For this reason, to facilitate comparison, the NetworkX library was used to produce a visualisation closer to the molecular structure. A comparison of the results obtained from the GNNExplainer vs. our HGE on Apigenin is shown in figure \ref{fig:GNNExplainerVsHGE}.

On the left side of the image, the molecular graph built from the results obtained by GNNExplainer is shown. In order to increase readability, carbon atoms have been coloured in light blue and oxygen atoms in red; the arcs have also been coloured using a different colour scheme according to the result obtained by the explainer. Specifically, non-relevant arcs were highlighted in grey, while relevant arcs were highlighted in red. As can be seen, GNNExplainer detects as relevant the two arcs connecting C-5/C-6/O-16 and the arc connecting C-12/C-13. This result shows only the above-mentioned molecular moieties as relevant for classification, while not recognising the other molecular moieties. In contrast, HGE provides a more comprehensive description of the molecular moieties involved in Apigenin bioactivity classification, detecting both the atoms relevant and their chemical species by labelling the pharmacophoric features associated with each substructure.

\subsection{Discussion}\label{sec4}
Experimental results show clearly that HGE performs very well in real applications with respect to both the prediction task and the explanation of the molecular moieties.
The main strength of the presented approach is the combination of different explanation layers (from single atoms to complex structures). Each layer is capable of catching different aspects of the molecule under investigation, and each step enriches the others, giving a complete interpretation of the molecular moieties useful to be classified as active for a specific target. As demonstrated in Figures \ref{fig:multitarget-apigenin}, 
\ref{fig:multitarget-trilaciclib} and \ref{fig:xai_CDK6_JAK2}, the moieties are predicted consistently across similar chemotypes for the annotated target, and the HGE prediction is consistent and robust even for drugs with bioactivity on distinct targets. As showed, in the Fig.2, the algorithm consistently identifies the aromatic regions and hydrogen bond acceptor features of the molecule, highlighting the same recognition patterns across the functional groups crucial for protein-ligand binding process. In Fig.3, compared to the previous case (Fig.2) where the recognition pattern identified the same moieties with the same importance, for Trilaciclib on CDK2 and CDK6 the algorithm slightly differentiated the role of the H-bond acceptor in N10 and O29. Probably, this difference is due to different potency magnitude of pharmacological effects of Trilacilib on CDK6 and CDK2.  The molecule is indeed known for its selectivity in binding the ATP-binding cleft of CDK6 and the potency on CDK2 is lower and its effect could be indirect. As a matter of fact, HGE identifies chemical structures compatible with protein binding process, without considering pharmacological effect or drug potency.
Figure 4 shows an example of how the algorithm is capable of retrieving and similarly interpreting very close chemotypes. In detail,  Pacritinib and SB 1578 show how molecule size do not influence the algorithm capability in mantaining the functional group recognition consistency.

\section{Conclusion}\label{sec5}

In this work the HGE framework has been proposed that combines different Grad-CAM explainability layers arranged in a message passing hierarchy to detect the molecular moieties involved in predicting the activity of a compound versus a given protein target.
To achieve our result, 20 GCNN classifiers have been trained on two different data sets derived from CheMBL to predict ligands' activity for 20 Kinase proteins. Our approach proved to be very effective in classification and was compared with Chemprop in this respect. Moreover, the moieties explanation ability was assessed using a suitable test set of 143 approved drugs derived from DrugBank, which are active on the 20 targets and comparing them with a proper docking analysis carried out by expert computational chemists.

Our experiments produced very good results with respect to classification and the moieties detection is highly precise. Such results set the framework as a very useful tool for Virtual Screening where our approach unveils potential hidden molecular moieties useful for binding a specific target. Finally, HGE is able to associate a given set of moieties with a particular target. As a consequence, very interesting insights can also be gained in the direction of drug repurposing if we use HGE to analyze a known drug that is known to own some of the moieties related to such a target.  HGE performance is shown to be good and reliable both in predicting the same moieties for a target (figure \ref{fig:xai_CDK6_JAK2}) and in identifying the same molecular features of the same molecule on different targets (figure \ref{fig:multitarget-apigenin}, \ref{fig:multitarget-trilaciclib}). These two complementary aspects demonstrate, in our opinion, a good reliability of HGE in identifying the important chemical moieties for small molecule-target recognition. In addition, the predictive comparison with GNNExplainer in Figure \ref{fig:GNNExplainerVsHGE} shows the different ability of our explainer in labelling information in purely chemical terms.
Future research will be dedicated to studying the ability of HGE to re-propose drugs and to understanding the link between classification properties and explainer prediction for assessing how to increase the amount of information that can be obtained from the explainer.

\section{Data and Code Availability}
Source code and is freely available at \url{https://github.com/CHILab1/HGE.git}. The data used for the test and train can be downloaded free of charge from the zenodo repository at the following link \url{https://zenodo.org/records/11125467?token=eyJhbGciOiJIUzUxMiJ9.eyJpZCI6ImY3ZDM0MjBiLWQyYjEtNGJiMi05YmY4LTE3Y2ZhNWRmMjVhMCIsImRhdGEiOnt9LCJyYW5kb20iOiJjYWQwNDdiODVjMzRmYTNkNjNhZjg5MTY3MTQxMGI5MSJ9.KzdfJ_C2_3kHrZ1tbbuz5xeRRETi_kGMMsYq2_EG2-46drEClcjlQzVHdMWIBL5pFbBjWBh4P94Em0M1qyD4dw}

\section{Supplementary information}
Supplementary file S1 gives a detailed description of the data used to construct the EMBER data set. In addition, there is a table showing the results obtained from the test with drugs active on multiple targets. 
Supplementary file S2 contains the results obtained from the HGE prediction for all the selected drugs, while supplementary file S3 contains the HGE plots of the drugs with activity on multiple targets reported to compare their robustness in the prediction of the molecular scaffold. 

\section*{Abbrevations}
\begin{itemize}
    \item CADD. Computer-Aided Drug Design
    \item VS. Virtual Screening
    \item SAR. Structure Activity Relationship
    \item XAI. Explainable AI
    \item EMBER. EMBedding multiplE molecular fingeRprints
    \item HGE. Hierarchical Graph Explainer
    \item IC50. Half Maximal Inhibitory Concentration
    \item I.C. Inactive Compound
    \item I.O. Inactive Oversampled
    \item GC Graph Convolution
    \item GCNN Graph Convolution Neural Network
    \item GAP Global Average Pooling
    \item Ki. Inhibition constant
    \item Kd. Dissociation constant
    \item MLP. Multi Layer Perceptron
    \item PDB. Protein Data Bank 
    \item ReLU. Rectified Linear Unit
    \item wBCE. Weighted Binary Cross-Entropy
    \item CAM Class Activation Map
    \item Grad-CAM. Gradient-weighted Class Activaction Mapping
    \item TP. True Positive
    \item P. Positive
    \item TN. True Negative
    \item N. Negative
    \item TP/P. True Positive vs Positive ratio
    \item AUC Area Under the Curve
    \item EF. Enrichment Factor
\end{itemize}

\section{Acknowledgments}
This work has been partially supported by project:\\
1.``SAMOTHRACE'' (Sicilian MicronanoTech Research And Innovation Center), cup project B73C22000810001, project code ECS\_00000022.\\
2. ``DARE'' (DigitAl lifelong pRevEntion), cup project B53C22006450001, project code PNC\_0000002.

\bibliographystyle{plain}
\bibliography{mybibfile}

\begin{thebibliography}{10}

\bibitem{schroeding_2023}
Schrodinger release 2024-1: Ligprep,.
\newblock Schrödinger, LLC, New York, NY, 2024.

\bibitem{Abdolmaleki_2017}
Azizeh Abdolmaleki, Jahan B.~Ghasemi, and Fatemeh Ghasemi.
\newblock Computer aided drug design for multi-target drug design: Sar /qsar, molecular docking and pharmacophore methods.
\newblock {\em Current Drug Targets}, 18(5):556–575, April 2017.

\bibitem{bahi2018}
Meriem Bahi and Mohamed Batouche.
\newblock Deep learning for ligand-based virtual screening in drug discovery.
\newblock In {\em 2018 3rd International Conference on Pattern Analysis and Intelligent Systems (PAIS)}, pages 1--5, 2018.

\bibitem{Bongini_Bianchini_Scarselli_2021}
Pietro Bongini, Monica Bianchini, and Franco Scarselli.
\newblock Molecular generative graph neural networks for drug discovery.
\newblock {\em Neurocomputing}, 450:242–252, August 2021.

\bibitem{carpenter2018}
Kristy~A Carpenter, David~S Cohen, Juliet~T Jarrell, and Xudong Huang.
\newblock Deep learning and virtual drug screening.
\newblock {\em Future Medicinal Chemistry}, 10(21):2557--2567, 2018.
\newblock PMID: 30288997.

\bibitem{Chen_Wang_2024}
Roufen Chen, Yuchen Wang, Zheyuan Shen, Chenyi Ye, Yu~Guo, Yan Lu, Jianjun Ding, Xiaowu Dong, Donghang Xu, and Xiaoli Zheng.
\newblock Discovery of potent csk inhibitors through integrated virtual screening and molecular dynamic simulation.
\newblock {\em Archiv Der Pharmazie}, 357(9):e2400066, September 2024.

\bibitem{Chen_Liu_Wu_2019}
Xin Chen, Xien Liu, and Ji~Wu.
\newblock Drug-drug interaction prediction with graph representation learning.
\newblock In {\em 2019 IEEE International Conference on Bioinformatics and Biomedicine (BIBM)}, page 354–361, November 2019.

\bibitem{chen2020concept}
Zhi Chen, Yijie Bei, and Cynthia Rudin.
\newblock Concept whitening for interpretable image recognition.
\newblock {\em Nature Machine Intelligence}, 2(12):772--782, 2020.

\bibitem{Datta_2024}
Shrayasi Datta, Chinmoy Ghosh, and J.~Pal Choudhury.
\newblock Classification of imbalanced datasets utilizing the synthetic minority oversampling method in conjunction with several machine learning techniques.
\newblock {\em Iran Journal of Computer Science}, September 2024.

\bibitem{Dick_Cocklin_2020}
Alexej Dick and Simon Cocklin.
\newblock Bioisosteric replacement as a tool in anti-hiv drug design.
\newblock {\em Pharmaceuticals}, 13(33):36, March 2020.

\bibitem{gilmer2017neural}
Justin Gilmer, Samuel~S Schoenholz, Patrick~F Riley, Oriol Vinyals, and George~E Dahl.
\newblock Neural message passing for quantum chemistry.
\newblock In {\em International conference on machine learning}, pages 1263--1272. PMLR, 2017.

\bibitem{Gimeno_Ojeda-Montes_Tomás-Hernández_Cereto-Massagué_Beltrán-Debón_Mulero_Pujadas_Garcia-Vallvé_2019}
Aleix Gimeno, María~José Ojeda-Montes, Sarah Tomás-Hernández, Adrià Cereto-Massagué, Raúl Beltrán-Debón, Miquel Mulero, Gerard Pujadas, and Santiago Garcia-Vallvé.
\newblock The light and dark sides of virtual screening: What is there to know?
\newblock {\em International Journal of Molecular Sciences}, 20(6):1375, March 2019.

\bibitem{Hareharen_2024}
K.~Hareharen, T.~Panneerselvam, and R.~Raj~Mohan.
\newblock Improving the performance of machine learning model predicting phase and crystal structure of high entropy alloys by the synthetic minority oversampling technique.
\newblock {\em Journal of Alloys and Compounds}, 991:174494, July 2024.

\bibitem{Jiang_Wu_Hsieh_Chen_Liao_Wang_Shen_Cao_Wu_Hou_2021}
Dejun Jiang, Zhenxing Wu, Chang-Yu Hsieh, Guangyong Chen, Ben Liao, Zhe Wang, Chao Shen, Dongsheng Cao, Jian Wu, and Tingjun Hou.
\newblock Could graph neural networks learn better molecular representation for drug discovery? a comparison study of descriptor-based and graph-based models.
\newblock {\em Journal of Cheminformatics}, 13(1):12, February 2021.

\bibitem{Kearnes_McCloskey_2016}
Steven Kearnes, Kevin McCloskey, Marc Berndl, Vijay Pande, and Patrick Riley.
\newblock Molecular graph convolutions: moving beyond fingerprints.
\newblock {\em Journal of Computer-Aided Molecular Design}, 30(8):595–608, August 2016.

\bibitem{Keseru_Makara_2006}
György~M. Keserű and Gergely~M. Makara.
\newblock Hit discovery and hit-to-lead approaches.
\newblock {\em Drug Discovery Today}, 11(15):741–748, August 2006.

\bibitem{Kimber_2021}
Talia~B. Kimber, Yonghui Chen, and Andrea Volkamer.
\newblock Deep learning in virtual screening: Recent applications and developments.
\newblock {\em International Journal of Molecular Sciences}, 22(9):4435, April 2021.

\bibitem{kipf2016semi}
Thomas~N Kipf and Max Welling.
\newblock Semi-supervised classification with graph convolutional networks.
\newblock {\em arXiv preprint arXiv:1609.02907}, 2016.

\bibitem{Landrum_2010}
Gregory Landrum.
\newblock Rdkit: Open-source cheminformatics;, 2010.

\bibitem{imbalanced_learn}
Guillaume Lema{{\^i}}tre, Fernando Nogueira, and Christos~K. Aridas.
\newblock Imbalanced-learn: A python toolbox to tackle the curse of imbalanced datasets in machine learning.
\newblock {\em Journal of Machine Learning Research}, 18(17):1--5, 2017.

\bibitem{Lin_MacKerell_2019}
Fang-Yu Lin and Alexander~D. MacKerell.
\newblock {\em Force Fields for Small Molecules}, volume 2022 of {\em Methods in Molecular Biology}, page 21–54.
\newblock Springer New York, New York, NY, 2019.

\bibitem{mendolia_2022}
Isabella Mendolia, Salvatore Contino, Giada De~Simone, Ugo Perricone, and Roberto Pirrone.
\newblock Ember—embedding multiple molecular fingerprints for virtual screening.
\newblock {\em International Journal of Molecular Sciences}, 23(44):2156, Jan 2022.

\bibitem{morris2019weisfeiler}
Christopher Morris, Martin Ritzert, Matthias Fey, William~L Hamilton, Jan~Eric Lenssen, Gaurav Rattan, and Martin Grohe.
\newblock Weisfeiler and leman go neural: Higher-order graph neural networks.
\newblock In {\em Proceedings of the AAAI conference on artificial intelligence}, volume~33, pages 4602--4609, 2019.

\bibitem{Pope2019ExplainabilityMF}
Phillip~E. Pope, Soheil Kolouri, Mohammad Rostami, Charles~E. Martin, and Heiko Hoffmann.
\newblock Explainability methods for graph convolutional neural networks.
\newblock {\em 2019 IEEE/CVF Conference on Computer Vision and Pattern Recognition (CVPR)}, pages 10764--10773, 2019.

\bibitem{Proietti_2024}
Michela Proietti, Alessio Ragno, Biagio~La Rosa, Rino Ragno, and Roberto Capobianco.
\newblock Explainable ai in drug discovery: self-interpretable graph neural network for molecular property prediction using concept whitening.
\newblock {\em Machine Learning}, 113(4):2013–2044, April 2024.

\bibitem{Rodríguez_Pérez_Bajorath_2020_2}
Raquel Rodríguez-Pérez and Jürgen Bajorath.
\newblock Interpretation of compound activity predictions from complex machine learning models using local approximations and shapley values.
\newblock {\em Journal of Medicinal Chemistry}, 63(16):8761–8777, August 2020.

\bibitem{Rodríguez_Pérez_Bajorath_2020_1}
Raquel Rodríguez-Pérez and Jürgen Bajorath.
\newblock Interpretation of machine learning models using shapley values: application to compound potency and multi-target activity predictions.
\newblock {\em Journal of Computer-Aided Molecular Design}, 34(10):1013–1026, October 2020.

\bibitem{ronneberger2015u}
Olaf Ronneberger, Philipp Fischer, and Thomas Brox.
\newblock U-net: Convolutional networks for biomedical image segmentation.
\newblock In {\em Medical Image Computing and Computer-Assisted Intervention--MICCAI 2015: 18th International Conference, Munich, Germany, October 5-9, 2015, Proceedings, Part III 18}, pages 234--241. Springer, 2015.

\bibitem{Santos_2017}
Rita Santos, Oleg Ursu, Anna Gaulton, A.~Patrícia Bento, Ramesh~S. Donadi, Cristian~G. Bologa, Anneli Karlsson, Bissan Al-Lazikani, Anne Hersey, Tudor~I. Oprea, and John~P. Overington.
\newblock A comprehensive map of molecular drug targets.
\newblock {\em Nature Reviews Drug Discovery}, 16(1):19–34, January 2017.

\bibitem{scarselligori2009}
Franco Scarselli, Marco Gori, Ah~Chung Tsoi, Markus Hagenbuchner, and Gabriele Monfardini.
\newblock The graph neural network model.
\newblock {\em IEEE Transactions on Neural Networks}, 20(1):61--80, 2009.

\bibitem{Selvaraju_2020}
Ramprasaath~R. Selvaraju, Michael Cogswell, Abhishek Das, Ramakrishna Vedantam, Devi Parikh, and Dhruv Batra.
\newblock Grad-cam: Visual explanations from deep networks via gradient-based localization.
\newblock {\em International Journal of Computer Vision}, 128(2):336–359, Feb 2020.
\newblock arXiv:1610.02391 [cs].

\bibitem{Sun_Huggins_2022}
Shan Sun and David~J. Huggins.
\newblock Assessing the effect of forcefield parameter sets on the accuracy of relative binding free energy calculations.
\newblock {\em Frontiers in Molecular Biosciences}, 9:972162, September 2022.

\bibitem{Unterthiner2015DeepLA}
Thomas Unterthiner, Andreas Mayr, and J{\"o}rg~Kurt Wegner.
\newblock Deep learning as an opportunity in virtual screening.
\newblock 2015.

\bibitem{Wang_Du_Song_2022}
Shudong Wang, Zhenzhen Du, Mao Ding, Alfonso Rodriguez-Paton, and Tao Song.
\newblock Kg-dti: a knowledge graph based deep learning method for drug-target interaction predictions and alzheimer’s disease drug repositions.
\newblock {\em Applied Intelligence}, 52(1):846–857, January 2022.

\bibitem{Wieder_Kohlbacher_Kuenemann_Garon_Ducrot_Seidel_Langer_2020}
Oliver Wieder, Stefan Kohlbacher, Mélaine Kuenemann, Arthur Garon, Pierre Ducrot, Thomas Seidel, and Thierry Langer.
\newblock A compact review of molecular property prediction with graph neural networks.
\newblock {\em Drug Discovery Today: Technologies}, 37:1–12, December 2020.

\bibitem{Xiong_Xiong_Chen_Jiang_Zheng_2021}
Jiacheng Xiong, Zhaoping Xiong, Kaixian Chen, Hualiang Jiang, and Mingyue Zheng.
\newblock Graph neural networks for automated de novo drug design.
\newblock {\em Drug Discovery Today}, 26(6):1382–1393, June 2021.

\bibitem{yang2019}
Kevin Yang, Kyle Swanson, Wengong Jin, Connor Coley, Philipp Eiden, Hua Gao, Angel Guzman-Perez, Timothy Hopper, Brian Kelley, Miriam Mathea, Andrew Palmer, Volker Settels, Tommi Jaakkola, Klavs Jensen, and Regina Barzilay.
\newblock Analyzing learned molecular representations for property prediction.
\newblock {\em Journal of Chemical Information and Modeling}, 59(8):3370--3388, 2019.
\newblock PMID: 31361484.

\bibitem{Ying_Bourgeois_2019}
Zhitao Ying, Dylan Bourgeois, Jiaxuan You, Marinka Zitnik, and Jure Leskovec.
\newblock Gnnexplainer: Generating explanations for graph neural networks.
\newblock {\em Advances in Neural Information Processing Systems}, 32, 2019.

\bibitem{ying2019gnnexplainer}
Zhitao Ying, Dylan Bourgeois, Jiaxuan You, Marinka Zitnik, and Jure Leskovec.
\newblock Gnnexplainer: Generating explanations for graph neural networks.
\newblock {\em Advances in neural information processing systems}, 32, 2019.

\bibitem{10.1093/nar/gkad1004}
Barbara Zdrazil, Eloy Felix, Fiona Hunter, Emma~J Manners, James Blackshaw, Sybilla Corbett, Marleen de~Veij, Harris Ioannidis, David~Mendez Lopez, Juan~F Mosquera, Maria~Paula Magarinos, Nicolas Bosc, Ricardo Arcila, Tevfik Kizilören, Anna Gaulton, A~Patrícia Bento, Melissa~F Adasme, Peter Monecke, Gregory~A Landrum, and Andrew~R Leach.
\newblock {The ChEMBL Database in 2023: a drug discovery platform spanning multiple bioactivity data types and time periods}.
\newblock {\em Nucleic Acids Research}, 52(D1):D1180--D1192, 11 2023.

\bibitem{Zeng_Tu_Liu_2022}
Xiangxiang Zeng, Xinqi Tu, Yuansheng Liu, Xiangzheng Fu, and Yansen Su.
\newblock Toward better drug discovery with knowledge graph.
\newblock 72:114–126, February 2022.

\bibitem{Zhang_Ren_2025}
Jungan Zhang, Yixin Ren, Yun Teng, Han Wu, Jingsu Xue, Lulu Chen, Xiaoyue Song, Yan Li, Ying Zhou, Zongran Pang, and Hao Wang.
\newblock Discovery of novel prmt1 inhibitors: a combined approach using ai classification model and traditional virtual screening.
\newblock {\em Frontiers in Chemistry}, 13, January 2025.

\bibitem{Zheng_2024}
Si~Zheng, Yaowen Gu, Yuzhen Gu, Yelin Zhao, Liang Li, Min Wang, Rui Jiang, Xia Yu, Ting Chen, and Jiao Li.
\newblock Machine learning-enabled virtual screening indicates the anti-tuberculosis activity of aldoxorubicin and quarfloxin with verification by molecular docking, molecular dynamics simulations, and biological evaluations.
\newblock {\em Briefings in Bioinformatics}, 26(1):bbae696, November 2024.

\bibitem{ZHOU202057}
Jie Zhou, Ganqu Cui, Shengding Hu, Zhengyan Zhang, Cheng Yang, Zhiyuan Liu, Lifeng Wang, Changcheng Li, and Maosong Sun.
\newblock Graph neural networks: A review of methods and applications.
\newblock {\em AI Open}, 1:57--81, 2020.

\end{thebibliography}

\end{document}